\documentclass[%
 reprint,
superscriptaddress,
bibnotes,
 amsmath,amssymb,
 aps,prl,
]{revtex4-2}

\usepackage{graphicx}
\usepackage{dcolumn}
\usepackage{bm}
\usepackage{xcolor}

\begin{document}

\preprint{}

\title{Nature of Sub-diffusion Crossover in Molecular and Polymeric Glass-Formers}

\author{H. Srinivasan}
\affiliation{Solid State Physics Division, Bhabha Atomic Research Centre, Mumbai 400085, India}
\affiliation{Homi Bhabha National Institute, Anushaktinagar, Mumbai 400094, India}
 
\author{V. K. Sharma}
\affiliation{Solid State Physics Division, Bhabha Atomic Research Centre, Mumbai 400085, India}
\affiliation{Homi Bhabha National Institute, Anushaktinagar, Mumbai 400094, India}

\author{V. García Sakai}
\affiliation{ISIS Neutron and Muon Centre, Rutherford Appleton Laboratory, Didcot, UK}

\author{S. Mitra}
\affiliation{Solid State Physics Division, Bhabha Atomic Research Centre, Mumbai 400085, India}
\affiliation{Homi Bhabha National Institute, Anushaktinagar, Mumbai 400094, India}

\begin{abstract}
A crossover from a non-Gaussian to Gaussian sub-diffusion has been observed ubiquitously in various polymeric/molecular glass-formers. We have developed a framework which generalizes the fractional Brownian motion (fBm) model to incorporate non-Gaussian features by introducing a jump kernel. We illustrate that the non-Gaussian fractional Brownian motion (nGfBm) model accurately characterizes the sub-diffusion crossover. From the solutions of the nGfBm model, we gain insights into the nature of van-Hove self-correlation in non-Gaussian subdiffusive regime, which are found to exhibit exponential tails, providing first such experimental evidence in molecular glass-formers. The validity of the model is supported by comparison with incoherent quasielastic neutron scattering data obtained from several molecular and polymeric glass-formers.

\end{abstract}

\maketitle

The problem of diffusion mechanism in supercooled liquids/glasses remains an essential component in the theory of glass transitions \cite{gotze1992,kob1997,debenedetti2001,arbe2002}. The diffusion processes in these media categorically violate the basic tenets of Brownian motion such as Fickianity (linear time-dependence of mean-squared displacement (MSD)) and Gaussianity (displacement distribution is Gaussian). However, such violations are often sensitive to the length and time scales explored and therefore inevitably lead to crossovers in diffusion mechanisms. These crossovers have now been experimentally observed in a wide range of systems including colloids \cite{raffaele2021,raffaele2022,wang2009,wang2012}, molecular/ionic liquids \cite{busselez2011, kofu2018}, polymers \cite{arbe2002,arbe2003,capponi2009,arbe1998}. Colloidal suspensions exhibit a crossover from non-Fickian to Fickian regime while yet remaining non-Gaussian\cite{raffaele2021,raffaele2022,wang2009,wang2012}. On the other hand, molecular and polymeric glass formers ubiquitously display a crossover from non-Gaussian to Gaussian diffusion while yet being non-Fickian (subdiffusive) \cite{busselez2011,kofu2018,arbe2002,arbe2003,capponi2009,arbe1998}. While the latter has been observed and investigated through various simulations\cite{arbe2002, busselez2011,arbe2002pre} and experiments\cite{arbe2002,busselez2011,kofu2018,arbe2003,capponi2009,arbe2002pre}, a first principles model has not yet been devised, which has precluded achieving a comprehensive understanding of the underlying basis for non-Gaussianity across these crossovers. Glass-formers exhibit a distinct exponential decay in their displacement distribution\cite{wang2009,slater2014,pinaki2007}, a feature observed in various complex fluids such as colloidal suspensions \cite{raffaele2022, raffaele2021, wang2009}, Si atoms in a silica melt\cite{pinaki2007,berthier2007jcp} and in Lennard-Jones particles\cite{pinaki2007,berthier2007jpcm}. This behavior is now understood to be a result of large deviations and  randomisation of number of jumps in particle displacement \cite{eli2020}. However, the precise nature of displacement distribution in glass-formers undergoing sub-diffusion crossover has not been investigated yet.

In this letter, we develop a model for non-Gaussian fractional Brownian motion (nGfBm) and show the emergence of sub-diffusion crossover through it. Further, our model also demonstrates that in the non-Gaussian sub-diffusive regime, displacement distribution clearly exhibits an exponential tail. We use the nGfBm model to analyze incoherent quasielastic neutron scattering (IQENS) data of molecular and polymeric glass-formers including ethylene glycol (EG) and its deep eutectic solvents, as well as MD simulation data of pure EG. Our findings provide the first experimental evidence of exponential tails in the displacement distribution of molecular glass-formers and also demonstrate the applicability of the model to other systems studied in literature\cite{busselez2011, kofu2018}.

The diffusion mechanisms and the crossovers therein can be characterized by investigating van Hove self-correlation function $G_s(r,t)$ or it's time Fourier transform, self-intermediate scattering function (SISF), $I_s(Q,t)$. In glass-formers the SISF is typically found to follow a stretched exponential function, $I_s (Q,t)=\exp[-{(t/\tau_s (Q))}^\beta ]$, where the stretching parameter $\beta$ characterizes the deviation from an exponential relaxation profile and $\tau_s$ is the characteristic relaxation time. Various experimental \cite{arbe2002, kofu2018, arbe2002pre, capponi2009, arbe1998} and computational\cite{busselez2011, arbe2002pre, capponi2009} studies show that polymer and molecular glass-formers exhibit a crossover in $\tau_s$ vs $Q$ relationship near the first maximum, $Q_0$ of the structure factor. Precisely speaking, for $Q < Q_0$, $\tau(Q) \sim Q^{-2/\beta}$, while for $Q > Q_0$, $\tau(Q) \sim Q^{-2}$. At low-Q values ($< Q_0$), juxtaposing the relationship $\tau_s  \sim Q^{-2/\beta}$ with the stretched exponential decay, inevitably leads to a Gaussian sub-diffusion with a MSD, $\left<\delta r^2 (t)\right> \sim t^\beta$. However, in the high $Q$ regime ($> Q_0$), where $\tau_s  \sim Q^{-2}$, the diffusion mechanism cannot be described within the Gaussian approximation \cite{arbe2002, busselez2011, arbe1998}. Therefore, this crossover in the behaviour of relaxation time, has been attributed to a transition from Gaussian (for $Q < Q_0$) to a non-Gaussian (for $Q > Q_0$) sub-diffusion in these media\cite{arbe2002, busselez2011, kofu2018}. In what follows, we present a characterization of the sub-diffusion crossover by formulating a Fokker-Planck equation for the nGfBm model. Our results demonstrate that the emergence of the sub-diffusion crossover is solely due to the non-locality induced by jump-kernel in the model.

The two salient features of glass-formers are the sub-diffusive behaviour due to strong memory effects and non-Gaussianity arising out of large jumps. To develop a physical model to describe a crossover from non-Gaussian to Gaussian sub-diffusion, it is essential to decouple these two features. Sub-diffusive phenomena that originate due to system’s strong memory can be modelled using the framework of fractional Brownian motion (fBm). The standard fBm, $B_\alpha (t)$, is a self-similar centered Gaussian process with an autocorrelation $\left< B_\alpha (t_1) B_\alpha (t_2) \right> = \left(t_1^\alpha + t_2^\alpha -|t_1 - t_2|^\alpha \right)$, where $\alpha \in (0,2)$. We define the particle displacements in these glassy media to be driven by the standard fBm, according to, $dx(t) = \sqrt{2 D_\alpha} dB_\alpha (t)$, $(0<\alpha<1)$, where $D_\alpha$ is the fractional diffusion constant with dimensions $\text{m}^2/\text{s}^\alpha$. Using the recent developments on calculus for a fBm-driven process \cite{duncan2000}, the Fokker-Planck equation for particle displacements can be obtained to be, $\frac{\partial G_s(x,t)}{\partial t} = \alpha t^{\alpha -1} D_\alpha \frac{\partial^2 G_s(x,t)}{\partial x^2}$ \cite{SM}. The solutions of this equation\cite{wei2022} for the initial condition, $G_s(x,0)=\delta(x)$, yields $G_s(x,t) = \left(4 \pi D_\alpha t^\alpha \right)^{-1} \exp \left[-x^2/ (4D_\alpha t^\alpha)\right]$. This equation describes a Gaussian sub-diffusion process where the MSD is $2 D_\alpha t^\alpha$ and all the higher cumulants of displacements are zero. In order to incorporate the non-Gaussian features in the standard fBm model, we propose the following equation for non-Gaussian fBm (nGfBm),
\begin{equation}\label{ngfbm1d}
    \frac{\partial G_s(x,t)}{\partial t} = \alpha t^{\alpha -1} \int_{-\infty}^\infty  dx'  \Lambda(x-x') \frac{\partial^2 G_s(x',t)}{\partial x'^2}
\end{equation}
where $\Lambda(x-x')$ is the jump-kernel which contains information about the spatially non-local nature of the diffusion process, allowing to account for large amplitude jumps which are the main source of non-Gaussianity in glass dynamics\cite{bier2008,vorselaars2007}. Using $\Lambda(x) = D_\alpha\delta(x)$ allows for only local or infinitesimally small displacements and therefore reproduces the standard fBm process. The general solutions to eq. \eqref{ngfbm1d} are readily obtained in the Fourier space using the SISF, $I_s(k,t) = I_0(k) \exp\left[ -k^2 \Lambda(k) t^\alpha\right]$, where $\Lambda(k) $ and $I_0(k)$ are the Fourier transforms of $\Lambda(x) $ and $G_s(x,0)$, respectively. 

The general form of jump-kernel in our study should exhibit transient non-Gaussian effects at short distances, and revert to Gaussian behaviour at long distances. To be more precise, we consider a characteristic length scale for jump processes, $x_0$. The jump-kernel induces non-Gaussian behavior for ($x \ll x_0$) and smoothly transitions to the Gaussian regime for ($x \gg x_0$). In the Fourier domain, these conditions can be simplified into two specific limiting rules: $\Lambda(k) \xrightarrow{kx_0\rightarrow 0} (x_0^2/\tau_j^\alpha)$ and $\Lambda(k)\xrightarrow{kx_0\rightarrow \infty} (k^{-2}/\tau_j^\alpha)$ \cite{SM}. Combining these rules, we propose the general jump-kernel as a series expansion
\begin{equation}
    \label{general jump-kernel}
    \Lambda(k) = \frac{k^{-2}}{\tau_j^\alpha}\sum_{n=1}^\infty c_n (kx_0)^{2n}
\end{equation}
where only the even terms are considered owing to the isotropic nature of the diffusion problem. The choice of the coefficients $\{c_n\}$ is dictated by conditions that the sum converges to unity for $kx_0 \rightarrow \infty$ and is proportional to $x_0^2$ for $kx_0 \rightarrow 0$. Various choices $\{c_n\}$ and their respective jump-kernels are listed in Table S1 \cite{SM}. Notable choices include $(-1)^{n-1}$ and $(-1)^{n-1}/n!$ which correspond to symmetric exponential and Gaussian jump-kernels are also referred to Model A and B in the list. Comparing the SISF of nGfBm from eq. \eqref{ngfbm1d} with the $e^{-(t/\tau(k))^\alpha}$ we get $\tau(k) \sim \tau_j (k^{2} \Lambda(k) )^{-1/\alpha}$. Fig. 1 shows the behaviour of the dispersion relationship for five different jump-kernels (Model A$-$E) listed in Table S1 (refer to \cite{SM}), showing a clear evidence of transition from Gaussian to non-Gaussian behaviour.

The exponential nature of displacement distribution emerges from the limiting behaviour $\Lambda(k)\xrightarrow{kx_0\rightarrow \infty} (k^{-2}/\tau_j^\alpha)$. The inverse Fourier transform in this limit is $\Lambda_h(x) \sim \frac{x_0}{2\tau_j^\alpha}e^{-|x|/x_0}$. It can be explicitly shown that this limiting form of jump-kernel leads to an exponential behaviour in the displacement distribution \cite{SM}. The log plot of displacement distributions, $G_s(x,t)$, at $t=\tau_j$ for different models shown in Fig. 1(b) bears evidence in support of the exponential nature. In the diminishing limit of $x_0$ and $\tau_j$, the jump-kernel reduces to $D_\alpha \delta(x)$ (where $D_\alpha = c_1 x_0^2/\tau_j^\alpha$) effectively restoring the Gaussian regime following fBm. Further, it is also notable for a given set of parameters ($x_0, \tau_j$), the displacement distribution approaches the Gaussian limit for $t \gg \tau_j$ irrespective of the choice of jump-kernel (Fig. S1). 

An exact analytical solution for the symmetric exponential kernel is derived to provide insights into real-space solutions, motivated by its reliability across various systems and its emergence as a limiting distribution in highly heterogeneous scenarios. For the exponential decay kernel considered in eq.\eqref{ngfbm1d} $I_s(k,t) = I_0(k) \exp\left[ \frac{-1}{1+(kx_0)^{-2}} \left[\frac{t}{\tau_j}\right]^\alpha \right]$. The exact solution of $I_s(k,t)$, obtained by inverting the Fourier transform \cite{SM}
\begin{equation}\label{1dsoln}
\begin{split}
G_s (x,t) =  e^{ - \hat t^\alpha  } \Bigg[ \frac{e^{ - \hat x}}
{{2x_0 }} \sum\limits_{n = 0}^\infty  {\frac{\hat t^{\alpha (n + 1) }}
{m_n} \hat x^n g_n \left(\hat x\right)}    + \delta (x) \Bigg]  \otimes G_0 
\end{split}
\end{equation}
where $G_0(x)$ is the initial condition (inverse Fourier transform of $I_0(k)$) of the system and $g_n(z)$ are functions governed by a recursive relationship $g_n(z) = z^{-1}(2n-1)g_{n-1}(z)+g_{n-2}(z)$, with the first two terms being $g_0(z)=1$ and $g_1(z)=z^{-1}(z+1)$. Here, $\hat x = (|x|/x_0)$ and $\hat t = (t/\tau_j)$ and $m_n = 2^n n!(n+1)!$. The solutions of the equation can also be given in terms of a series of modified Bessel functions, $K_n(z)$\cite{SM}. In the strongly heterogeneous limit ($|x|/x_0 \rightarrow 0$) the solutions given in eq.\eqref{1dsoln}  shows an exponential decay in the leading term, reminiscing the universal behavior of displacement distribution in glass-forming systems\cite{pinaki2007,eli2020}; and the corrections to the exponential behaviour is governed by the algebraic power series in eq.\eqref{1dsoln}. The contribution of these corrections are also related to the ratios $(t/\tau_j)$ in the series. The appearance of exponential tails is more prominent when the system is probed at times much smaller than $\tau_j$. As the system approaches glass-transition, the effective frequency of jumps decreases, leading to sharper display of exponential tails. This is consistent with the consensus that the exponential behaviour of displacement distribution is enhanced near the glass-transition\cite{pinaki2007}. In general, our model tends to exhibit stronger affinity to exponential tails with increasing heterogeneity ($x_0$ and $\tau_j$) \cite{SM}.
\begin{figure}
    \centering
    \includegraphics[width=1\linewidth]{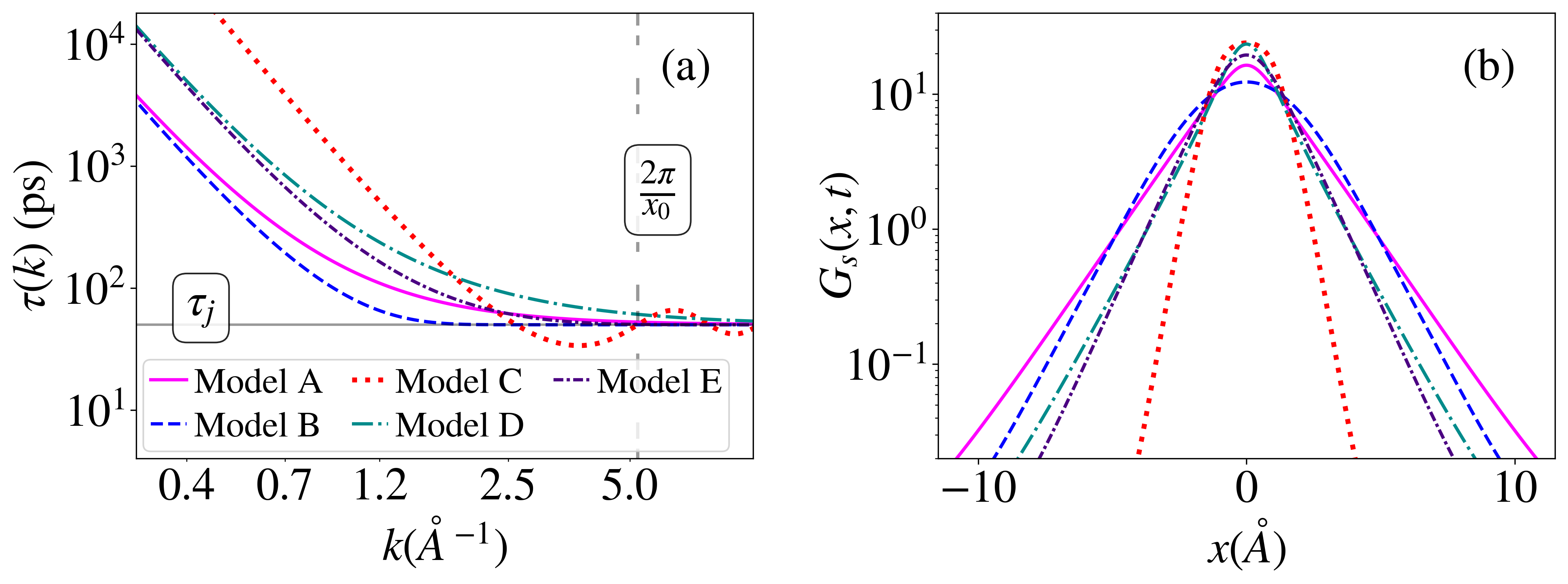}
    \caption{(a) Variation of relaxation time with momentum transfer $k$ for different jump-kernel models listed in Table S1 using the same of set of parameters ($x_0, \tau_j, \alpha$). The horizontal (solid) and vertical (dashed) lines represent values of $\tau_j$ and $2\pi/x_0$, respectively. (b) The calculated displacement distribution, $G_s(x,t)$ at $t = \tau_j$, for the case of each of these jump-kernels. The final distributions are convoluted with a Gaussian ensemble for $G_s(x,0)$.}
    \label{jump models}
\end{figure}

As observed in various other glass-forming systems \cite{arbe1998, arbe2002, capponi2009, kofu2018, busselez2011}, the IQENS data of ethylene glycol (EG) and its associated deep eutectic solvents (DESs) - EG + $\text{ZnCl}_2$ (1:4 molar ratio) and EG + LiCl (1:3 molar ratio) follow a stretched exponential relaxation profile based on characteristic timescale $\tau_s$ and stretching exponent $\beta$ (described in SM \cite{SM}). Liquid ethylene glycol (EG) is known to exhibit stretched exponential relaxation \cite{crupi2003, sobolev2007}, and it is anticipated that DESs based on EG will also display this characteristic, given their resemblance to supercooled liquids\cite{swarup2020}. The average relaxation time $\tau_a(Q) = \tau_s(Q)\beta^{-1}\Gamma(\beta^{-1})$ is calculated from IQENS data fitting. Fig. S4 clearly indicate the crossover from Gaussian dynamics at low $Q$ to non-Gaussian behaviour at higher $Q$-values for EG and the DESs.

\begin{figure}
\centering
\includegraphics[width=\linewidth]{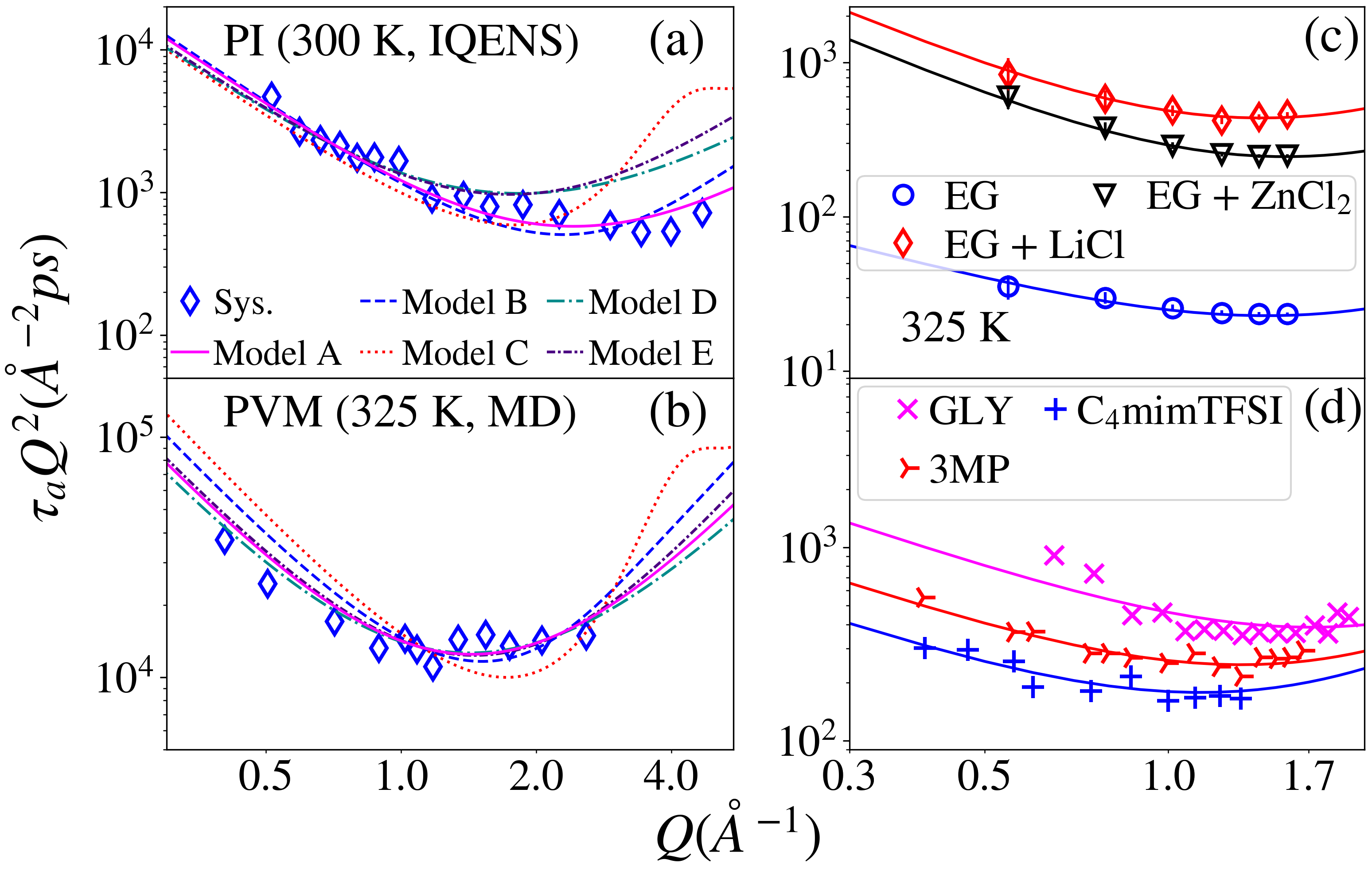}
\caption{\label{tauavg}(a) Plot of $\tau_a Q^2$ vs $Q$ for (a) polyisoprene (PI) \cite{arbe2002} and (b) poly vinyl methyl ether (PVM) \cite{capponi2009} with fits based on five different jump-kernels (Models A-E). The exponential kernel (Model A), suits the best over a wide $Q$-range and hence employed in modelling systems with limited $Q$-range. Plot of $\tau_a Q^2$ vs $Q$ for (c) EG, EG + LiCl and EG + $\text{ZnCl}_2$ at 325 K and (d) different systems from literature (GLY: glycerol\cite{busselez2011}, 3MP: 3-methylpentane\cite{kofu2018}, $\text{C}_4$mimTFSI: 1-butyl-3-methylimidazolium bis(trifluoromethanesolfonyl)imide\cite{kofu2018}) along with fits based on Model A.}
\end{figure}

Fig. \ref{tauavg} (a)-(b) show the variation of relaxation time $\tau_a$ in polymeric systems, as observed from IQENS experiments for polyisoprene (PI) \cite{arbe2002} and MD simulations of polyvinyl methylether (PVM) \cite{capponi2009}. Similar plots for the systems investigated in this study (EG and their DESs) and other molecular organic glass-formers\cite{kofu2018, busselez2011} are also shown in Fig. \ref{tauavg} (c) and (d), respectively. As illustrated in the plots, these systems show a crossover from $Q^{-2/\beta}$ (for $Q < 1 \text{\AA}^{-1}$) to $Q^{-2}$ (for $Q > 1 \text{\AA}^{-1}$).

To model the observed crossover in the experimental IQENS studies, we extend the nGfBm (eq. \eqref{ngfbm1d}) to 3D systems using a three-dimensional fBm process with an autocorrelation of the form $\left< \mathbf{r} (t_1) .\mathbf{r}(t_2) = 3 D_\alpha (t_1^\alpha + t_2^\alpha -|t_1 -t_2|^\alpha \right>$. It has been shown that an n-dimensional fBm process can be constructed as a linear superposition of $n$ independent fBm processes \cite{jeon2010}. This allows us to construct an extension of eq. \eqref{ngfbm1d} for the 3D nGfBm \cite{SM}. The 3D jump-kernels can also be chosen based on the eq. \eqref{general jump-kernel} with parameters $r_0$ and $\tau_j$ subject to an additional constraint of radial symmetry. The $Q$-dependence of the average relaxation time follow $\tau_a(Q) = (Q^2\Lambda(Q))^{-1/\alpha}$, which we use to fit the $Q$-dependence of the measured $\tau_a$ for all the systems using models based on different jump-kernels listed in Table S1 \cite{SM}. 

For PI and PVM, Fig. \ref{tauavg} (a) and (b) demonstrate that model A (exponential jump-kernel) provides the fits in the extended $Q$-range. Fig. \ref{tauavg} (c)-(d) showcase the versatility of the exponential kernel \cite{diff_kernel} across a diverse range of systems, including EG, EG based DESs, glycerol (GLY), 3-methylpentane (3MP) and the ionic liquid $\text{C}_4$mimTFSI. This highlights the robustness of the nGfBm model which works for IQENS measurements carried out using different spectrometers with varying resolutions and in particular also emphasizes the larger applicability of exponential kernel. 

Evidently, the model represents the data very well and captures the crossover from Gaussian dynamics at low $Q$ to non-Gaussian behaviour at higher $Q$-values for all the systems. The model parameters $r_0$ and $\tau_j$ provide valuable insights into the extent of dynamical heterogeneity in the system. For the specific case of the exponential kernel, the crossover point in $Q$-space is linked to $r_0$ and can be precisely estimated as, $Q^* = \sqrt{(1/\beta) -1}/r_0$ ($\beta\ne 1$). This implies that crossover point is inversely related to the extent of spatial heterogeneity in the non-local diffusive process. As $r_0\rightarrow 0$, representing a completely homogeneous limit, $Q^*$ tends to infinity, resulting in Gaussian sub-diffusion regime at all length scales. The values of $Q^*$ for different glass-forming systems with varying fragility index are listed in Table S3 in the SM \cite{SM}. The ability to model self-diffusion in wide range of systems with different fragility indices indicates the robustness of the nGfBm model. Notably, $r_0$ shows minimal temperature variation, suggesting that the crossover point $Q^*$ is largely unaffected by temperature changes. Meanwhile, the jump time, $\tau_j$, shows a remarkable increase with decrease in temperature, as glass transition is approached. It is notable that the increase in the jump times $\tau_j$ essentially dictates that, the non-Gaussianity in the diffusion process is enhanced as the system approaches glass-transition. We also find that $Q^*$ typically lies in the 1 – 2 $\text{\AA}^{-1}$ (Table S4) [24] for all the glass-formers that have been investigated so far. These results, while consistent with the existing literature, reinforces the idea that the origin of sub-diffusion crossover is linked to the length scale of non-local jumps in the diffusion process rather than its relationship with the first maxima of the structure factor peak.

\begin{figure}
\centering
\includegraphics[width=0.77\linewidth]{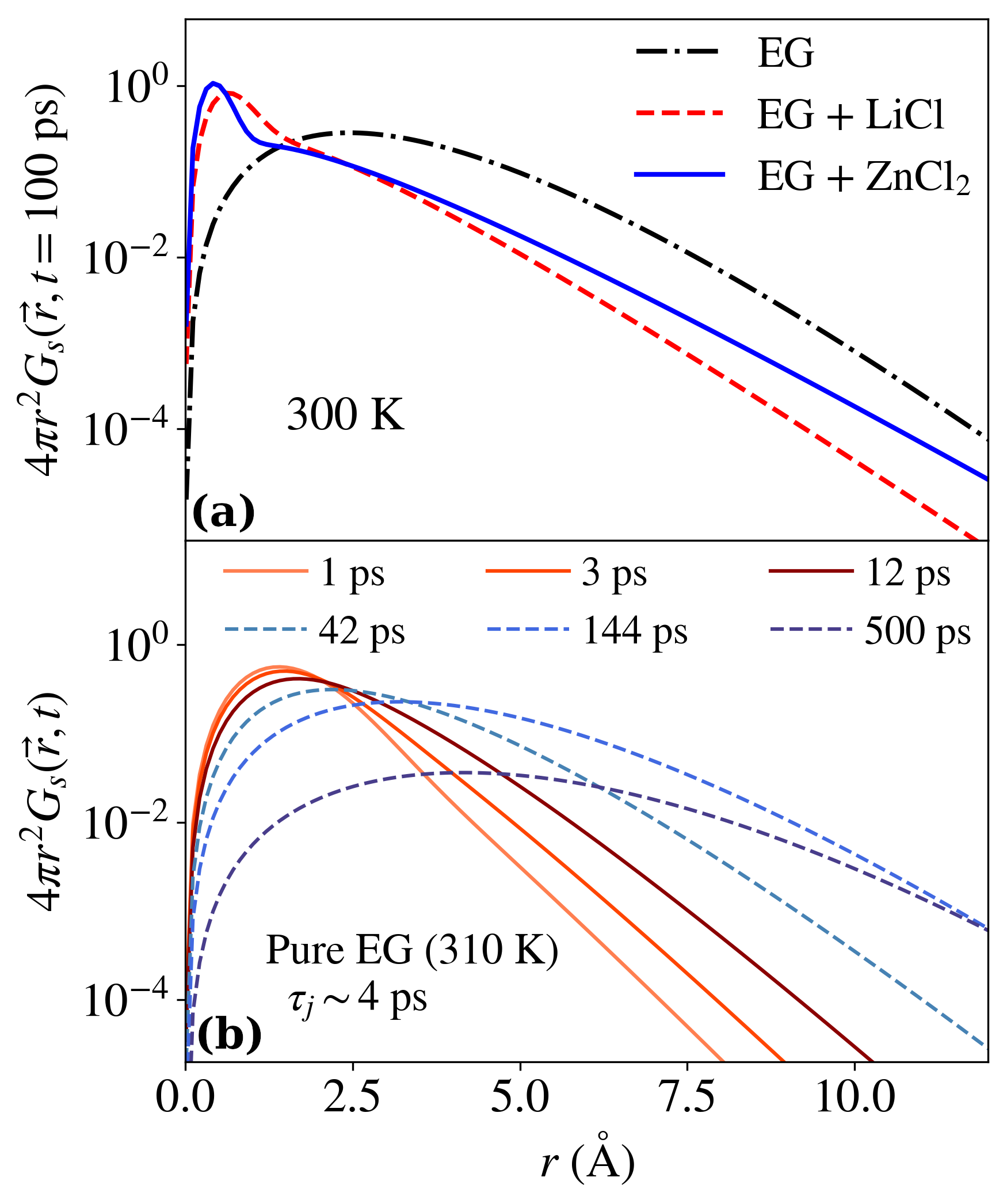}
\caption{\label{gsrt} (a) The radial van-Hove self-correlation function calculated from eq. \eqref{3dsoln} based on the parameters ($\tau_j, \beta, r_0, \left<u^2\right>$) extracted from experimental IQENS fits for pure EG and DESs (EG + LiCl, EG + $\text{ZnCl}_2$) at T = 300 K and $t$ = 100 ps. (b) Radial van-Hove self-correlation function for pure EG (310 K) calculated at different times $t$; $t > 10\tau_j$ are shown by broken lines and $t < 10\tau_j$ are shown by solid lines.}
\end{figure}

In order to shed more light on the transition from non-Gaussian to Gaussian behaviour, we calculate the 3D van-Hove self-correlation function from the model. Focussing on radially symmetric kernel, we have $I(Q,t) = \exp \left[ -(Qr_0)^2/(1+(Qr_0)^2) (t/\tau_j)^\beta\right]$. Using this, the van Hove self-correlation function in this case is, 
\begin{widetext}   
\begin{equation}\label{3dsoln}
G_s \left( {\mathbf{r},t} \right) = e^{ - \left( {t/\tau _j } \right)^\beta  } \left[ {\frac{1}
{{8\pi r_0^3 }}e^{ - \left( {\frac{r}
{{r_0 }}} \right)} \sum\limits_{n = 0}^\infty  {\frac{{(t/\tau _j )^{\beta \left( {n + 1} \right)} }}
{{(n + 1)!n!}}\left( {\frac{r}
{{2r_0 }}} \right)^{n - 1} g_{n - 1} \left( {\frac{r}
{{r_0 }}} \right)}  + \delta (\mathbf{r})} \right] \otimes \left(\frac{3}{4\pi \left<u^2\right>}\right)^{3/2} e^{-3r^2/(4\left< u^2 \right>)}
\end{equation}
\end{widetext}
The first term in eq. \eqref{3dsoln} describes the heterogeneous diffusion process with a typical exponential displacement distribution in the leading order, while the second term represents the Gaussian thermal cloud generated by fast localized dynamics. 
 For $t \ll \tau_j$, the radial van-Hove self-correlation function ($4\pi r^2 G_s (\mathbf{r},t)$ ) is typically a sum of Gaussian thermal cloud near origin with exponential tails at large values of $r$. As time progresses, the contribution of the diffusive component becomes dominant and the second term decreases at a rate governed by $e^{-(t/\tau_j)^\beta}$. Fig. \ref{gsrt}(a) shows the plots of $4\pi r^2 G_s (\mathbf{r} ,t=100 ps)$ for the DESs and pure EG, directly computed with set of parameters ($r_0,\tau_j,\alpha,\left< u^2\right>$ ) obtained from the fits of the QENS spectra, at 300 K. The significantly different values of $\tau_j$ for EG and DESs, being ~ 7 ps and ~ 200 ps, respectively, account for the marked disparity in their curves. In DESs, longer $\tau_j$ causes the initial peaks ($r < 2 \text{\AA}$) in van-Hove self-correlation function to be more pronounced. These peaks have vanished in the case of pure EG, as the fast local dynamics have completely relaxed and the dynamics is described as a purely diffusive process at 100 ps. Further, the tails prominently exhibit an exponential decay in the cases of both the DESs but shows a nearly Gaussian behaviour for pure EG. A clearer perspective of these changes are apparent from the plots of radial van-Hove self-correlation of pure EG (310 K) at different times (ranging from 1 to 500 ps) in Fig. \ref{gsrt}(b). Typically, it is observed that the exponential tails are discernible for $t  < 10 \tau_j$. The curves for $t < 10 \tau_j$ and $t > 10 \tau_j$ are marked by solid and broken lines respectively to clearly indicate this feature and exhibit the crossover from non-Gaussian to Gaussian regime.

In order to check the veracity of the model, we have calculated the radial van-Hove self-correlation function from MD simulations of pure EG and tried to fit it with eq. \ref{3dsoln}. The parameters from the fit were optimized so as to describe the simulated data over a reasonable time-window (0-500 ps). In order to achieve a consistent model, the fitting was carried out at different times and model parameters were optimized across all the times. The extracted parameters are found to be fairly consistent with experimental observations. The simulated van-Hove self-correlation function and their respective fits are given in Fig. S8. 

The diffusion landscape in glass-formers exhibits various crossovers entailing different mechanisms. Universally, molecular and polymeric glass-formers exhibit an inherent crossover from non-Gaussian to Gaussian sub-diffusion. In this work, we have characterized the nature of particle displacements across this crossover by augmenting the framework of fBm to incorporate non-local jumps. The constructed model provides a generalized mathematical structure that can be used to describe a number of such processes that simultaneously involve both non-Markovianity and spatial jumps in a diffusion process. We present a generalized jump-kernel that elucidates the subdiffusion crossover, revealing the emergence of an exponential kernel as a consequence of physical constraints. The case of exponentially distributed jumps is solved providing a link between the crossover point and the extent of heterogeneity in the system. Notably, this framework also highlights the exponential nature of van-Hove self-correlation function during the non-Gaussian subdiffusion regime.
\nocite{mandelbrot1968}
\nocite{iris}
\nocite{kumar2013}
\nocite{gygli2020}
\nocite{abbott2007}
\nocite{namd}
\nocite{charmm2012}

\bibliography{MS_refs}

\begin{thebibliography}{36}
\bibitem [{SM()}]{SM}%
  \BibitemOpen
  \href@noop {} {\bibinfo {title} {{S}ee {S}upplemental {M}aterial for detailed
  mathematical derivations and more information about experiments and
  simulations, which also contains refs. [30-36]}}\BibitemShut {NoStop}%
\end{thebibliography}

\end{document}